\documentclass[prb,reprint]{revtex4-1}

\usepackage{graphicx}
\usepackage{bm}% bold math
\usepackage{hyperref}% add hypertext capabilities
\usepackage{color}
\begin{document}

\title {Phase Diagrams for the $\nu$ = 1/2 Fractional Quantum Hall Effect in Electron Systems Confined to Symmetric, Wide GaAs Quantum Wells}

\author{J.~Shabani, Yang Liu, M.~Shayegan, L. N. Pfeiffer, K. W. West, and K. W. Baldwin}

\affiliation{Department of Electrical Engineering, Princeton University, Princeton, NJ 08544, USA}

\date{\today}

\begin{abstract}

  We report an experimental investigation of fractional
  quantum Hall effect (FQHE) at the even-denominator Landau level filling factor
  $\nu$ = 1/2 in very high quality wide GaAs quantum wells, and at
  very high magnetic fields up to 45 T. The quasi-two-dimensional
  electron systems we study are confined to GaAs quantum wells with
  widths $W$ ranging from 41 to 96 nm and have variable densities in the range of
$\simeq 4 \times 10^{11}$ to $\simeq 4 \times 10^{10}$ cm$^{-2}$. We present several experimental phase diagrams for the stability of the
  $\nu=1/2$ FQHE in these quantum wells. In general, for a given $W$, the 1/2 FQHE is stable in a limited range of intermediate densities where it has a bilayer-like charge distribution; it makes a transition to a compressible phase at low densities and to an insulating phase at high densities. The densities at which the
  $\nu=1/2$ FQHE is stable are larger for narrower quantum wells. Moreover, even a slight charge distribution asymmetry destabilizes the $\nu=1/2$ FQHE and turns the electron system into a compressible state. We also present a plot of the symmetric-to-antisymmetric subband separation ($\Delta_{SAS}$), which characterizes the inter-layer tunneling, vs density for various $W$. This plot reveals that $\Delta_{SAS}$ at the boundary between the compressible and FQHE phases increases \textit{linearly} with density for all the samples. There is no theoretical explanation for such a simple dependence. Finally, we summarize the experimental data in a diagram that takes into account the relative strengths of the inter-layer and intra-layer Coulomb interactions and $\Delta_{SAS}$. We conclude that, consistent with
  the conclusions of some of the previous studies, the $\nu=1/2$ FQHE observed in
  wide GaAs quantum wells with symmetric charge distribution is stabilized by a delicate balance between the inter-layer and intra-layer interactions, and is very likely described by a two-component ($\Psi_{311}$) state.

\end{abstract}

\pacs{}

\maketitle

\section{Introduction}

The fractional quantum Hall effect (FQHE) \cite {Tsui.PRL.1982} is predominantly seen in
high-quality two-dimensional (2D) electron systems in the lowest
($N=0$) Landau level at odd-denominator fillings $\nu$
\cite{Jain.CF.2007}. In the first, excited ($N=1$)
Landau level, a FQHE exists at the \textit{even-denominator} filling
$\nu=5/2$ \cite{Willett.PRL.1987, Pan.PRL.1999}. This enigmatic FQHE has
become the focus of considerable theoretical and experimental
attention, partly because of its potential application in topological
quantum computing \cite{Nayak.Rev.Mod.Phys.2008}. Despite numerous experimental
efforts during the past two decades, however, a thorough understanding
of its origin remains elusive. In particular, it is yet not known
whether or not the spin degree of freedom is necessary to stabilize
this state. If yes, then the 5/2 FQHE state could be described by a
two-component, Halperin-Laughlin ($\Psi_{331}$) wavefunction
\cite{Halperin.HPA.1983}. But if the 5/2 FQHE is stable in a fully
spin-polarized 2D electron system, then it is likely to be the
one-component, Moore-Read (Pfaffian) state \cite{Moore.Nuc.Phy.1991}. The latter is
of enormous interest as it is expected to obey non-Abelian statistics
and have use in topological quantum computing \cite{Nayak.Rev.Mod.Phys.2008}.

The possibility of an even-denominator FQHE in the {\it lowest} Landau
level, e.g. at $\nu = 1/2$, has been theoretically discussed in
numerous publications \cite{Halperin.HPA.1983, Rezayi.BAPS.1987,
  Chakraborty.PRL.1987, Yoshioka.PRB.1989, MacDonald.SS.1990,
  Moore.Nuc.Phy.1991, He.PRB.1991, Greiter.PRL.1991, He.PRB.1993,
  Halperin.SS.1994, Nomura.JPSJ.2004, Papic.PRB.2009a, Papic.PRB.2009, Storni.PRL.2010,
  Peterson.PRB.2010, Peterson.PRB.2010b}. Experimentally, FQHE states
at $\nu=1/2$ have been seen in electron systems confined to either
double \cite{Eisenstein.PRL.1992}, or wide \cite{Suen.PRL.1992,
  Suen.PRL.1992b, Suen.Thesis.1993, Suen.PRL.1994, Manoharan.PRL.1996,
  Shayegan.SST.1996, Shayegan.TALDS.1999, Luhman.PRL.2008,
  Shabani.PRL.2009a, Shabani.PRL.2009} GaAs quantum well (QW) systems; $\nu=1/2$ FQHE was also reported very recently in bilayer graphene system \cite{Ki.cond.mat.2013}. In wide GaAs QWs, FQHE has also been seen at other even-denominator fillings, namely at $\nu=3/2$ \cite{Suen.PRL.1994} and at $\nu = 1/4$ \cite{Luhman.PRL.2008, Shabani.PRL.2009a, Shabani.PRL.2009}. In a
double QW with negligible inter-layer tunneling but comparable inter-layer and intra-layer Coulomb interactions, it is generally
accepted that the $\nu=1/2$ FQHE is stabilized by the additional
(layer) degree of freedom, and is described by the two-component,
$\Psi_{331}$ state \cite{Rezayi.BAPS.1987, Chakraborty.PRL.1987,
  Yoshioka.PRB.1989, MacDonald.SS.1990, He.PRB.1991, He.PRB.1993,
  Halperin.SS.1994, Nomura.JPSJ.2004, Papic.PRB.2009a, Papic.PRB.2009, Storni.PRL.2010,
  Peterson.PRB.2010, Peterson.PRB.2010b}; in this case the components
are the layer indices. However, the situation is more subtle for the
case of electrons in a single, wide QW where the electron-electron repulsion lifts the potential energy near the well center and creates an effective barrier \cite{Suen.PRL.1992, Suen.PRL.1992b, Suen.Thesis.1993, Suen.PRL.1994, Manoharan.PRL.1996,
  Shayegan.SST.1996, Shayegan.TALDS.1999, Luhman.PRL.2008,
  Shabani.PRL.2009a, Shabani.PRL.2009, Suen.PRB.1991, Lay.PRB.1994}. Although the system can have a "bilayer-like" charge distribution at sufficiently high densities, the inter-layer tunneling, quantified by the symmetric-to-antisymmetric
subband separation ($\Delta_{SAS}$), can be substantial. Moreover, in a QW with fixed well-width, the magnitude of $\Delta_{SAS}$ can be tuned from small to large values by decreasing the electron density in the QW while keeping the total charge distribution symmetric ("balanced"). When $\Delta_{SAS}$ is negligible compared to the intra-layer Coulomb energy ($e^2/ 4 \pi \epsilon l_B$) then, similar to the double-QW system, $\Psi_{331}$ is the likely ground state if a $\nu=1/2$ FQHE is observed ($l_B =
\sqrt{\hbar/eB}$ is the magnetic length and $\epsilon$ is the
dielectric constant). If
$\Delta_{SAS}$ is a significant fraction of $e^2/ 4 \pi \epsilon l_B$, however, then it is likely that the
$\nu=1/2$ FQHE state is a one-component, Pfaffian state \cite{Greiter.PRL.1991}.

Here we present results of our extensive experimental study of the
$\nu=1/2$ FQHE in very high quality, wide GaAs QWs with well widths
($W$) ranging from 41 to 96 nm and tunable densities ($n$) in the range of
$\simeq 4 \times 10^{11}$ to $\simeq 4 \times 10^{10}$ cm$^{-2}$. Our
data, taken at low temperatures and very high perpendicular magnetic fields
($B$ up to 45 T) allow us to determine the most comprehensive set of
experimental conditions for the stability of the $\nu=1/2$ FQHE in
symmetric, wide GaAs QWs. We present our data in several experimental phase
diagrams, including a $d/l_{B}$ vs $\Delta_{SAS}/(e{^2}/ 4 \pi
\epsilon l_{B})$ diagram; $d/l_{B}$ is the ratio of the inter-layer distance ($d$) and the magnetic length, and is a measure of the relative strengths the intra-layer
($e^2/ 4 \pi \epsilon l_B$) and inter-layer ($e^2/ 4 \pi \epsilon d$)
interactions. Our transport data also reveal that, even in the narrowest QWs, making the charge distribution slightly asymmetric ("imbalanced"), the $\nu=1/2$ FQHE disappears and is replaced by a compressible state. We conclude that, consistent with the
conclusions of previous experimental \cite{Suen.Thesis.1993,
  Suen.PRL.1994, Shabani.PRL.2009a} and theoretical \cite{He.PRB.1993,
  Storni.PRL.2010, Peterson.PRB.2010} studies, the $\nu=1/2$ FQHE
observed in wide GaAs QWs with symmetric charge distribution is
likely a two-component state. Comparing our experimental $d/l_{B}$ vs
$\Delta_{SAS}/(e{^2}/ 4 \pi \epsilon l_{B})$ phase diagram to a recently
calculated diagram \cite{Peterson.PRB.2010}, we find that, while there is some
overall qualitative agreement, there are also significant
quantitative discrepancies.

\section{Experimental details}

Our samples were grown by molecular beam epitaxy and each consists of
a GaAs QW bounded on both sides by undoped Al$_x$Ga$_{1-x}$As
barrier layers ($x\simeq$ 0.24 to 0.30) and Si $\delta$-doped layers. The well widths of these
samples range from 41 to 96 nm, but the focus of our work is on
narrower QWs ($W<70$ nm) where $\Delta_{SAS}$ is large. These narrower QW
samples typically have low-temperature mobilities of 250 to 600
m$^{2}$/Vs and the Al composition in their barriers is $x\simeq$ 0.24. The wider QW samples used in older studies \cite{Suen.PRL.1992, Suen.PRL.1992b, Suen.Thesis.1993, Suen.PRL.1994} had mobilities of $\simeq$ 100 m$^{2}$/Vs and their barriers had $x\simeq$ 0.30. All the samples had a van der Pauw geometry, an approximately 3 mm $\times$ 3 mm square shape, and were fitted with an evaporated Ti/Au front-gate and a Ga
or an In back-gate to change the 2D density while keeping the charge
distribution in the well symmetric. We carried out measurements in
dilution refrigerators with base temperatures of $\simeq$ 30 mK,
housed in either a 16 T superconducting magnet, a 35 T resistive
magnet, or a 45 T hybrid (resistive plus superconducting) magnet.

Most of the samples used in our study were cut from GaAs wafers which were not rotated during the epitaxial growth. This resulted in a non-uniformity of the growth rate across the wafer surface and an uncertainty in the QW width. To determine the QW width more accurately, we carefully measured and analyzed the low-field Shubnikov-de Haas oscillations in each sample to obtain $\Delta_{SAS}$ as a function of $n$. We also performed calculations of the charge distribution and the QW's potential and energy levels by solving Schroedinger and Poisson's equations self-consistently (at zero magnetic field) while treating the QW width $W$ as an adjustable parameter. We then compared the measured $\Delta_{SAS}$ vs $n$ data to the calculations using $W$ as a fitting parameter. We show examples of these fits in Section V (Fig. 5) where we present the $\Delta_{SAS}$ vs $n$ data. The values of $W$ we quote here are from such fits, and have an estimated absolute accuracy of about $\pm$5\%, although their relative accuracy is better than about $\pm$2\%. The quoted $W$ also agree with the nominal QW widths based on the epitaxial growth rates to within about $\pm$10\%.\\

{\section{G\lowercase{a}A\lowercase{s} single wide quantum wells}

We first briefly describe the electron system confined to a
modulation-doped wide QW. When electrons at very low density are
placed in a single wide QW, they occupy the lowest electric subband and have
a single-layer-like (but rather thick in the growth direction) charge
distribution. As more electrons are added to the well while keeping
the charge distribution symmetric, their electrostatic repulsion
forces them to pile up near the QW boundaries and the electron charge
distribution appears increasingly bilayer-like \cite{Suen.PRL.1992, Suen.PRL.1992b, Suen.Thesis.1993, Suen.PRL.1994,
  Manoharan.PRL.1996, Shayegan.SST.1996, Shayegan.TALDS.1999,
  Luhman.PRL.2008, Shabani.PRL.2009a, Shabani.PRL.2009, Suen.PRB.1991, Lay.PRB.1994}. At high $n$
the electrons typically occupy the lowest two, symmetric and
antisymmetric electric subbands; these are separated in energy by
$\Delta_{SAS}$, which for a bilayer system can be considered as the
inter-layer tunneling energy. An example of the potential energy and the charge distribution in
such a system is given in Fig.~1 inset where we show the results of
our self-consistent calculations for $n=4.34 \times 10^{11}$ cm$^{-2}$
electrons symmetrically distributed in a 41-nm-wide QW.

A crucial property of the electron system in a wide QW is that both
$\Delta_{SAS}$ and $d$ (the inter-layer separation, see Fig. 1 inset),
which characterize the coupling between the layers, depend on density:
Increasing the density makes $d$ larger and $\Delta_{SAS}$ smaller so
that the system can essentially be tuned from a (thick)
single-layer-like electron system at low density to a bilayer one by
increasing the density \cite{Suen.PRL.1992,
  Suen.PRL.1992b, Suen.Thesis.1993, Suen.PRL.1994, Shayegan.SST.1996,
  Manoharan.PRL.1996, Shayegan.TALDS.1999, Luhman.PRL.2008,
  Shabani.PRL.2009a, Shabani.PRL.2009, Suen.PRB.1991, Lay.PRB.1994}. This evolution with density
plays a decisive role in the properties of the correlated electron
states in the system \cite{Suen.PRB.1991, Suen.PRL.1992,
  Suen.PRL.1992b, Suen.Thesis.1993, Suen.PRL.1994, Shayegan.SST.1996,
  Manoharan.PRL.1996, Shayegan.TALDS.1999, Luhman.PRL.2008,
  Shabani.PRL.2009a, Shabani.PRL.2009, Suen.PRB.1991, Lay.PRB.1994}. At low densities, when
$\Delta_{SAS}$ is large, the electron system exhibits odd-denominator
FQHE characteristic of a standard, single-layer sample. In particular,
at $\nu=1/2$ the electron system is compressible and there is no
FQHE. At very high densities, it shows an insulating phase near
$\nu=1/2$, which likely signals the formation of a bilayer Wigner
crystal \cite{Manoharan.PRL.1996, Shayegan.SST.1996, Shayegan.TALDS.1999}. And eventually, at the highest
densities, $\Delta_{SAS}$ becomes small and the system is very much
bilayer-like; it typically exhibits FQHE at odd-denominator fillings
which have $even$ $numerators$, consistent with the presence of two
layers in parallel. In the intermediate density regime, on the other
hand, a FQHE is seen at $\nu=1/2$ in the highest quality samples. This
FQHE, and its transition into a compressible state as the density is
lowered, are the subjects of our study presented here.

Experimentally, we control both the density and the charge
distribution symmetry in our samples via front- and back-side gates,
and by measuring the densities of the two occupied electric subbands
from Fourier transforms of the low-field Shubnikov-de Haas
magnetoresistance oscillations. These Fourier transforms typically
exhibit two peaks whose frequencies are directly proportional to the
subband densities. By carefully monitoring the evolution of these
frequencies as a function of density and, at a fixed density, as a
function of the values of the back and front gate biases, we can
determine and tune the symmetry of the charge distribution
\cite{Suen.Thesis.1993, Suen.PRL.1994, Shayegan.SST.1996,
  Manoharan.PRL.1996, Shayegan.TALDS.1999, Shabani.PRL.2009a,
  Shabani.PRL.2009, Suen.PRB.1991, Lay.PRB.1994}. Note that $\Delta_{SAS}$ is directly proportional
to the difference between the subband densities and, for a fixed
density in the QW, is smallest when the total charge distribution in
the QW is symmetric. The data presented here were all taken on wide
QWs with symmetric ("balanced") charge distributions, except in Fig. 4. Also, all the quoted values
of $\Delta_{SAS}$ are from the measured Shubnikov-de Haas oscillation
frequencies.

\begin{figure} \includegraphics[width =
  0.45\textwidth]{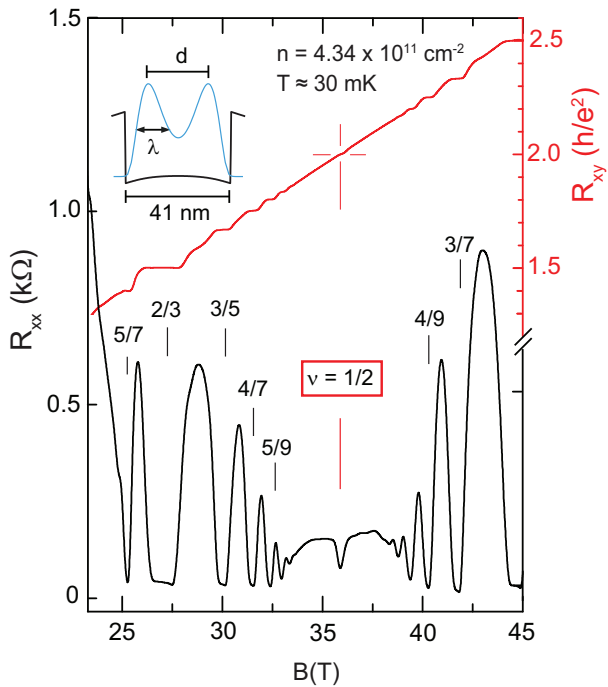} \caption{(color online) Longitudinal
    ($R_{xx}$) and Hall ($R_{xy}$) resistance data for a 41-nm-wide QW
    at a density of $4.34 \times 10^{11}$ cm$^{-2}$. The inset shows
    the electron distribution (blue curve) and potential energy (black curve)
    calculated self-consistently at zero magnetic
    field. The "inter-layer distance" $d$ and full-width-at-half maximum $\lambda$ of each "layer" are also indicated. \label{Fig1}}
\end{figure}

\begin{figure} \includegraphics[width =
  0.45\textwidth]{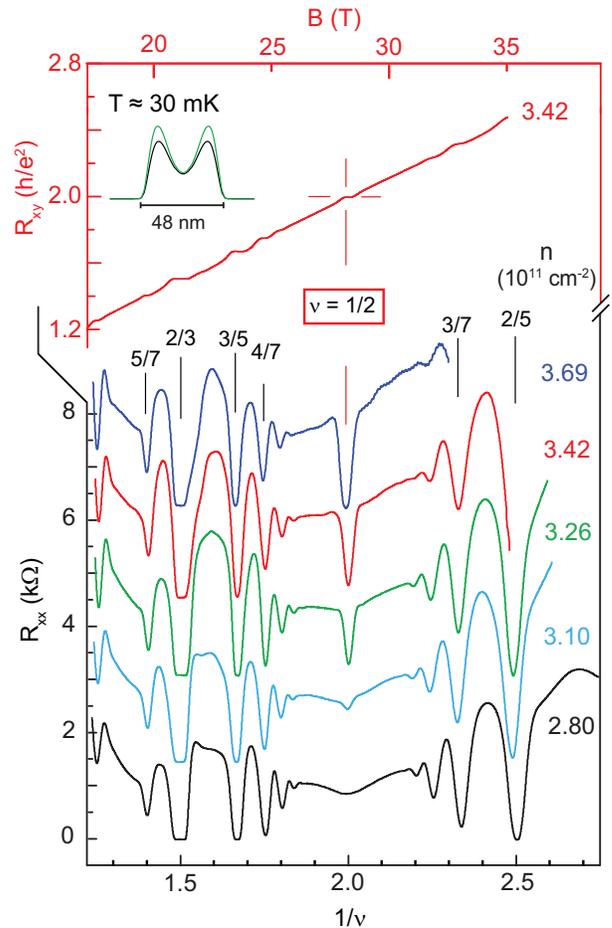} \caption{(color online) $R_{xx}$ plotted vs inverse filling
    factor for a 48-nm-wide QW for different densities. All traces
    were taken for symmetric total charge distributions ("balanced"
    QW). Except for the lowest density, the traces are offset vertically for clarity. In each trace, $R_{xx}\simeq$ 0 at $\nu=2/3$. Note that for the $n = 3.69\times 10^{11}$ cm$^{-2}$ (dark blue) trace, $R_{xx}$ at $\nu=1/2$ essentially vanishes. For the $n = 3.42\times 10^{11}$ cm$^{-2}$ (red) trace, in the top section of the figure we also show its corresponding $R_{xy}$ trace (in red), illustrating a
well-developed $R_{xy}$ plateau quantized at $2h/e^{2}$; the top
    axis is the perpendicular magnetic field for these two red traces. The inset shows the self-consistently calculated charge distributions for $n$ =
    2.80 (black) and 3.26 $\times 10^{11}$ cm$^{-2}$ (green). \label{Fig2}} \end{figure}

\begin{figure} \includegraphics[width =
  0.45\textwidth]{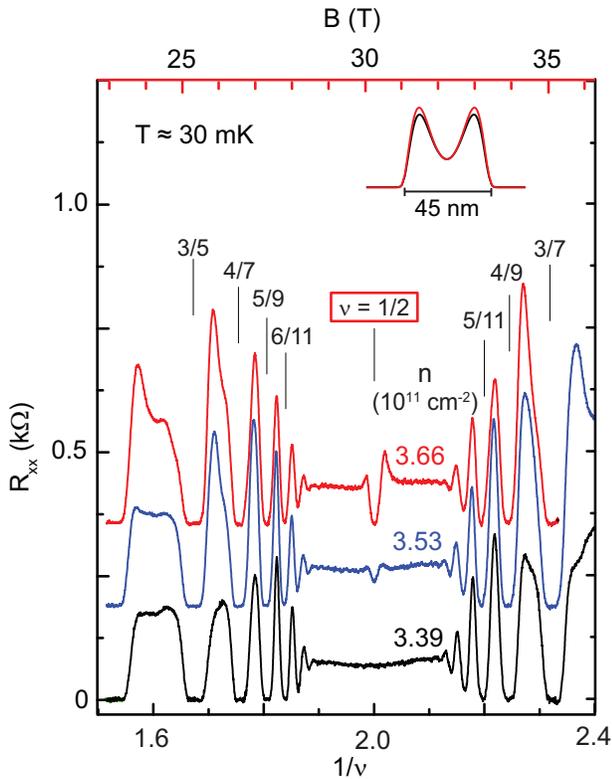} \caption{(color online) $R_{xx}$ plotted vs
    inverse filling factor for a 45-nm-wide QW for different
    densities while the total charge distribution was kept
    symmetric. The upper two traces are shifted vertically for clarity. The top
    axis is the perpendicular magnetic field for the $n = 3.66\times
    10^{11}$ cm$^{-2}$ trace. The inset shows the calculated charge distributions for $n$
    = 3.39 (black) and 3.66 $\times 10^{11}$ cm$^{-2}$ (red). \label{Fig3}} \end{figure}

\begin{figure}[htp]
\centering \includegraphics[width=0.45\textwidth]{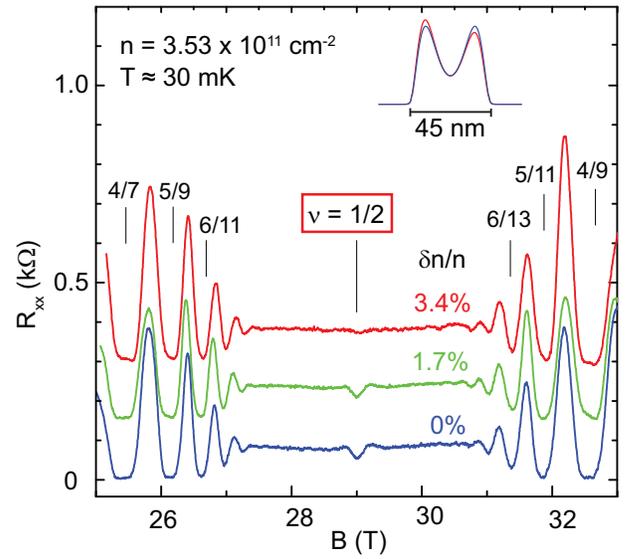} \caption{(color
  online) Evolution of the $\nu$=1/2 FQHE in the 45-nm-wide GaAs quantum well with total density of $n=3.53 \times 10^{11}$cm$^{-2}$ as the charge distribution is made asymmetric via applying front- and back-gate voltage biases. Inset: Self-consistently calculated electron distributions for $\delta n=0$ and $\delta n/n=0.034$. }
  \end{figure}

{\section{Examples of magnetoresistance data}

Figure 1 provides an example of the longitudinal ($R_{xx}$) and Hall ($R_{xy}$)
resistance traces at high magnetic fields for a 41-nm-wide QW at a
density of $4.34 \times 10^{11}$ cm$^{-2}$. $R_{xx}$ minima are observed at
numerous Landau level fillings such as $\nu=$ 2/3, 3/5, 4/7, 5/9, 6/11, 7/13, and 2/5, 3/7, 4/9, 5/11, 6/13, 7/15, 
attesting to the very high quality of the sample. These
odd-denominator FQHE states and their relative strengths resemble
those seen in standard, single-layer, 2D electrons confined to high
quality GaAs QWs with narrower well widths. Of particular interest here is of course the well-developed
FQHE at $\nu=1/2$ as evidenced by a deep $R_{xx}$ minimum and a
developing $R_{xy}$ plateau at $2h/e^{2}$. This FQHE has
no counterpart in standard 2D electron systems in narrow QWs. The
charge distribution and potential for this QW at $n = 4.34 \times
10^{11}$ cm$^{-2}$, calculated self-consistently (at $B=0$), are also
shown in Fig. 1 inset. The charge distribution is bilayer-like, with
$d$ denoting the inter-layer distance and $\lambda$ the
full-width-at-half-maximum of the charge distribution in each "layer".

In Fig.~\ref{Fig2} we show the density dependence of the $R_{xx}$
traces for a 48-nm-wide QW sample. The bottom
$x$-axis used in Fig. 2 is the inverse filling factor in order to
normalize the magnetic field for the different densities. At the
lowest density (bottom trace) the FQHE states are very similar to
those seen in the standard, narrow 2D electron systems; more
precisely, they are all at odd-denominator fillings and there is no
FQHE at $\nu=1/2$. As the density increases, the traces reveal a clear
sharp minimum developing in $R_{xx}$ at $\nu$ = 1/2 which is
accompanied by a quantized plateau in $R_{xy}$ at $2h/e^{2}$ (see the $R_{xy}$ trace in the top part of Fig. 1). In the
inset we show the calculated charge distributions for $n=2.80 \times
10^{11}$ cm$^{-2}$ (black), when the state at $\nu=1/2$ is
compressible, and $n=3.26 \times 10^{11}$ cm$^{-2}$ (green), when the
$\nu=1/2$ FQHE is observed. It is clear in Fig. 2 that a very small change in
the electron density is sufficient to turn the ground state at
$\nu=1/2$ from compressible to FQHE.

\begin{figure} \centering
\includegraphics[width =0.45 \textwidth]{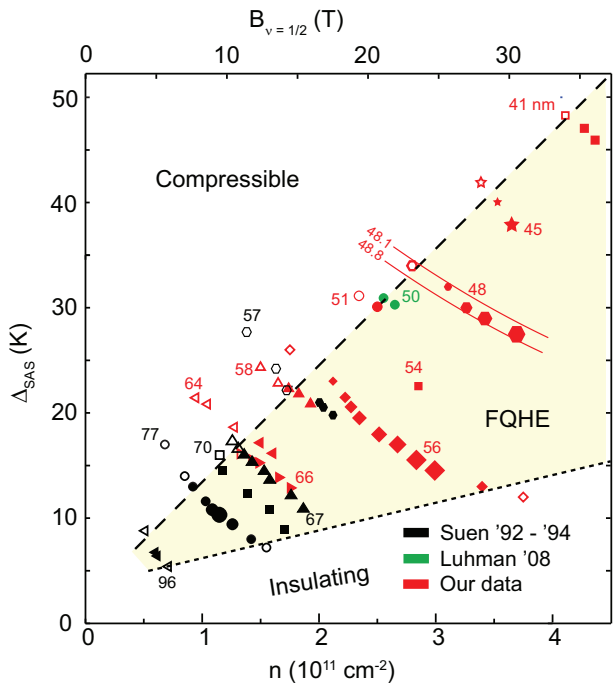}
\caption{(color online) Experimentally
    measured subband separation energy ($\Delta_{SAS}$) is plotted as a function of
    density ($n$) for electron systems with symmetric charge distributions confined to wide GaAs QWs. The accuracy of the measured $\Delta_{SAS}$ is about $\pm$2\%. The well widths are given (in units of nm) next to each set of data points, and are determined from fitting each set to our self-consistent (Hartree) calculations, examples of which are shown by thin red curves for the 48-nm-wide QW (see text). Data points in black are from Refs. \cite{Suen.PRL.1992, Suen.PRL.1992b, Suen.Thesis.1993, Suen.PRL.1994} and those in green from Ref. \cite{Luhman.PRL.2008}. Filled symbols represent the presence of a $\nu=$ 1/2 FQHE and open symbols its absence; the size of each filled symbol for the 45-, 48-, 56-, and 77-nm-wide QWs provides an estimate for the strength of the observed $\nu=$ 1/2 FQHE.
    The boundary between the FQHE and compressible states appears to
    be a $straight$ line (dashed line) over the entire range of QW
    widths and densities in our study.   \label{Fig5} } \end{figure}

\begin{figure}
  \centering
  \includegraphics[width=0.45\textwidth]{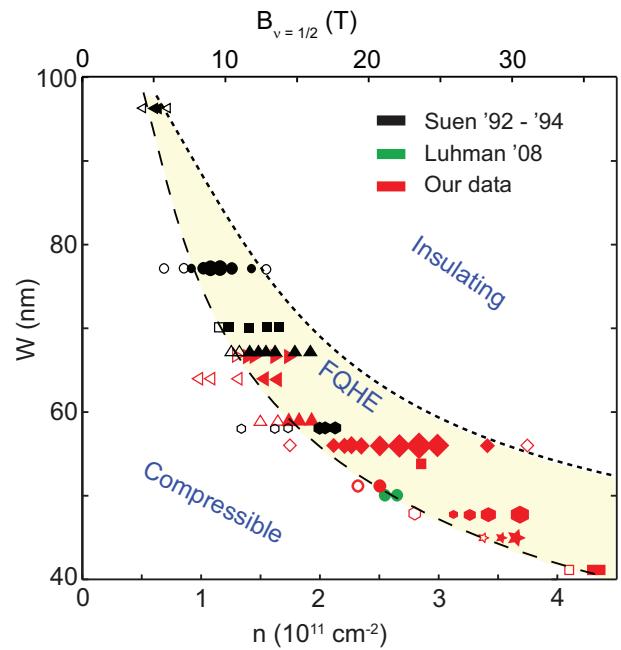}
  \caption{(color online) The well-width ($W$) vs density ($n$) phase diagram for
    the state of the electron system at $\nu=$ 1/2 in symmetric, wide GaAs
    QWs. The symbols have the same meaning as in Fig. 5. \label{fig:Fig4}}
\end{figure}

Similar data for a slightly narrower, 45-nm-wide QW are shown in
Fig. 3. The data again attest to the very high quality of this sample,
as they exhibit high-order FQHE at odd-denominator fillings up to 7/15
and 8/15. The trend for the ground state at $\nu=1/2$ is similar to
the 48-nm-wide sample of Fig. 2: At the lowest density the state is
compressible but as the density is slightly increased a FQHE
appears. But note in Fig. 3 that the density for the transition between the
compressible and FQHE ground state is somewhat larger for the 45-nm-wide sample
compared to the slightly wider QW of Fig. 2.

In Fig. 4 we demonstrate the crucial role of electron charge distribution symmetry on the stability of the $\nu=1/2$ FQHE.
Here we fix the density in the 45-nm-wide sample at $n=3.53\times 10^{11}$ cm$^{-2}$, and show $R_{xx}$ vs magnetic field measured for different QW charge distribution symmetries. We
change the symmetry by increasing the density by $\delta n$ via
applying the front-gate bias and decreasing the density by $\delta n$
via the back-gate bias}. The $\nu=1/2$ FQHE is strong when the QW is
symmetric ($\delta n/n=0$). It becomes weak when $\delta n/n = 0.017$ and is completely destroyed when $\delta n/n = 0.034$. This
evolution is also very similar to what is observed in electron systems confined
to wider GaAs QWs \cite{Suen.Thesis.1993, Suen.PRL.1994, Manoharan.PRL.1996, Shayegan.SST.1996, Shayegan.TALDS.1999, Shabani.PRL.2009a}. It shows that, even in relatively narrow QWs, the $\nu=1/2$ FQHE is destabilized by a slight asymmetry in the charge distribution. \\

\section{Phase diagrams for the stability of $\nu=1/2$ FQHE}

We have made measurements similar to those shown in Figs. 2 and 3 for
several samples with different QW widths, and summarize the results in
various "phase diagrams" shown in Figs. 5-7. In all these figures, the charge distribution in the wide QW is symmetric, and the  filled symbols indicate that the $\nu=1/2$ FQHE is stable. The size of the
filled symbols for data from some representative QW widths ($W=$ 45,
48, 56, and 77 nm) give an approximate indication of the strength of
the FQHE as deduced, e.g., from the depth of the $\nu=1/2$ $R_{xx}$
minimum or from the measured energy gaps \cite{Suen.Thesis.1993,
  Suen.PRL.1994}. The open symbols in Figs. 5-7 denote the absence of
a $\nu=1/2$ FQHE. In all the samples, the trend is the same: The
$\nu=1/2$ FQHE is seen in an intermediate density range which depends
on the QW width, but turns into a compressible state when the density
is sufficiently lowered. In Figs. 5-7 we mark the approximate boundary
between the FQHE and the compressible state with a dashed curve. This
boundary is a main focus of our work presented here. At sufficiently high densities, the electron system turns into an
insulating phase whose characteristics suggest the formation of a
pinned bilayer Wigner crystal \cite{Manoharan.PRL.1996, Shayegan.SST.1996, Shayegan.TALDS.1999}. We have
indicated the boundary between the FQHE and the insulating phase with
a dotted curve in Figs. 5-7. This boundary, and the properties of the
insulating phase are interesting in their own right, but are beyond
the scope of our study. We note, however, that this boundary is
difficult to determine in narrower QWs because of the very high
densities, and hence very high magnetic fields, that are required for its access.

\begin{figure}[htp]
  \centering \includegraphics[width=0.45 \textwidth]{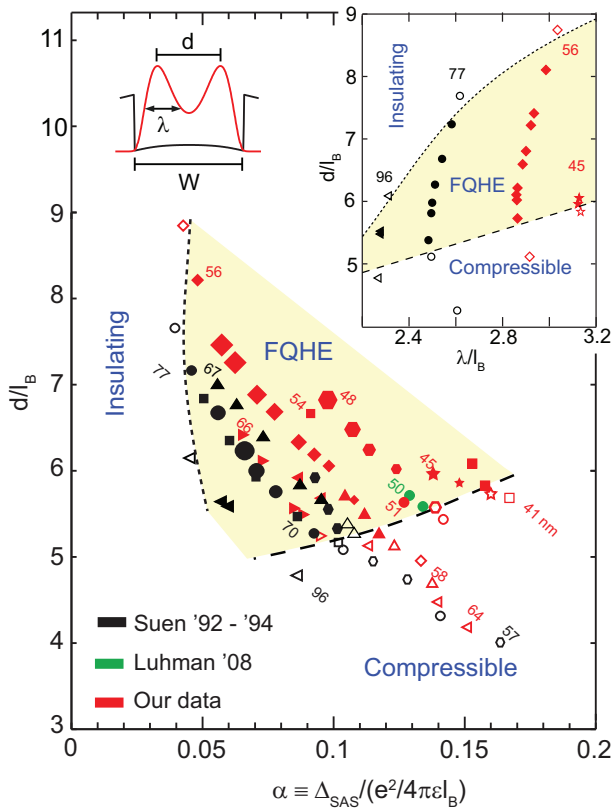} \caption{(color
    online) The $d/l_B$ vs $\alpha=\Delta_{SAS}/(e{^2}/ 4 \pi \epsilon
    l_{B})$ phase diagram for the observation of the $\nu$ = 1/2
    FQHE in symmetric, wide GaAs QWs. The well widths $W$ are given in units of nm. The upper-left inset shows
    a typical electron charge distribution
    calculated self-consistently; the "inter-layer distance" $d$ and full-width-at-half maximum $\lambda$ of each "layer" are also indicated. The upper-right inset is the $d/l_{B}$ vs
    $\lambda/l_{B}$ phase diagram; for clarity only data points for well widths $W$ = 45, 56, 77 and 96 nm are shown.} \end{figure}

Before discussing these phase diagrams, we would like to highlight some additional information that Fig. 5 provides: For a given well-width $W$, $\Delta_{SAS}$ decreases with increasing $n$, and this dependence allows us to determine reasonably precise values for $W$. This is important because, as stated in Section II, many of our samples were not rotated during the molecular beam epitaxial growth and $W$ is not precisely known. To determine $W$, we performed self-consistent (Hartree) calculations of the charge distribution and potential, and hence $\Delta_{SAS}$, while keeping $W$ as a fitting parameter. Examples of the results of such calculations are shown by two thin, red, solid lines in Fig. 5 for $W=$ 48.1 and 48.8 nm. It is clear that the measured data points for the sample whose $W$ we quote as 48 nm fall between these two lines. Using a similar procedure, we determined $W$ for all other samples, except for the sample of Ref. \cite {Luhman.PRL.2008} (green circles in Fig. 5). For this sample, a $W=$ 50 nm was quoted in Ref. \cite{Luhman.PRL.2008} but no measurements of $\Delta_{SAS}$ were reported. We thus used $W=$ 50 nm and in Fig. 5 we plot our calculated $\Delta_{SAS}$ for the two densities reported in Ref. \cite{Luhman.PRL.2008}. We would like to emphasize that, for consistency in our presentation, we used the same calculations to determine $W$ for the older samples of Suen $et$ $al.$ \cite{Suen.PRL.1992, Suen.PRL.1992b, Suen.Thesis.1993, Suen.PRL.1994} (black data points in Fig. 5). We have found that there is a small discrepancy between $W$ determined from our fits and those quoted previously. In particular, we find $W=$ 57, 67, and 70 nm while in Refs. \cite{Suen.PRL.1992, Suen.PRL.1992b, Suen.Thesis.1993, Suen.PRL.1994} the quoted values are 60, 68, and 71 nm, respectively. These discrepancies mainly stem from the differences in the self-consistent calculations and the band parameters used. Given the accuracies of the measured $\Delta_{SAS}$ and also the self-consistent calculations, we estimate the overall absolute accuracy of the quoted $W$ to be about $\pm$5\%. Their relative accuracy, however, is better than about $\pm$2\%. Our quoted $W$ also agree with the nominal QW widths based on the epitaxial growth rates to within about $\pm$10\%.

Returning to the phase diagrams in Figs. 5-7, each provides a different perspective on the
stability of the $\nu=1/2$ FQHE in wide GaAs QWs. Figure 5 plot is rather unique in that the parameters for both axes, $\Delta_{SAS}$ and density, are experimentally measured quantities. The plot clearly demonstrates that the $\nu=1/2$ FQHE is only stable in a range of intermediate densities which depends on the QW width. More remarkably, it reveals that the boundary between the FQHE and compressible ground states (dashed line in Fig. 5) appears to be well described by essentially a $straight$ line. We are not aware of any theoretical calculations which predict such a
simple (linear) boundary for the transition between the compressible and FQHE phases in wide QWs at $\nu=1/2$.

Figure 6 provides perhaps the simplest phase diagram as it gives at a glance the range of densities within which
the $\nu=1/2$ FQHE is stable for electrons confined to a symmetric GaAs QW of width $W$. It is clear that the narrower the QW, the higher the density range where the $\nu=1/2$ FQHE is seen. Figure 6 also indicates that in very wide QWs the FQHE is stable only in a very small range of very low densities.

Finally, we provide the phase diagram of Fig. 7 which is often used to
discuss the origin and stability of the $\nu=1/2$ FQHE in a wide QW or
in a double QW system \cite{He.PRB.1993, Suen.Thesis.1993, Suen.PRL.1994, Peterson.PRB.2010}. This diagram takes into account
the competition between three relevant energies: (i) the inter-layer tunneling
energy ($\Delta_{SAS}$), (ii) the intra-layer (in-plane) Coulomb
energy ($e{^2}/ 4 \pi \epsilon l_{B}$), and (iii) the inter-layer
Coulomb energy ($e^{2}/4 \pi \epsilon d$). The coordinates in the
diagram of Fig. 7 are the ratios of $\Delta_{SAS}/(e{^2}/ 4 \pi
\epsilon l_{B})$, which we denote as $\alpha$, and $(e^{2}/4 \pi
\epsilon l_{B})/ (e^{2} /4 \pi \epsilon d)$ which is equal to
$d/l_{B}$. It is worth noting that in a wide QW, as the density increases, $\alpha$
decreases because both $\Delta_{SAS}$ and $l_{B}$ (at $\nu=1/2$)
decrease, and $d$ increases.

Besides the above three energies, another potentially relevant energy is the Zeeman energy $E_{Z}=|g^*|\mu_{B}B$ where
  $g^*$ is the effective Lande g-factor and $\mu_{B}$ is the Bohr
  magneton. Using the GaAs band value of $g^*=-0.44$, $E_{Z}=0.3$
  K/T. This is smaller than $\Delta_{SAS}$ at magnetic fields where
  the $\nu=1/2$ FQHE is observed in our samples (see, e.g., Fig. 5),
  implying that the spin degree of freedom is important. However, in
  our wide GaAs QW samples, $g^*$ is typically enhanced by a factor of
  $\simeq 10$ or larger \cite{Liu.PRB.2011}, so that $E_Z$
  is more than twice larger than $\Delta_{SAS}$. We therefore expect
  the spin degree of freedom not to play a role in the stability of
  the $\nu=1/2$ FQHE in our samples. This is consistent with data in
  tilted magnetic field data which show that, starting near the
  compressible-FQHE boundary, the $\nu=1/2$ FQHE gets stronger at
  small tilt angles \cite{Lay.PRB.1997, Luhman.PRL.2008}. Such
  strengthening is consistent with $\Delta_{SAS}$ becoming effectively
  smaller because of the parallel field component \cite{Hu.PRB.1992},
  thus moving the electron system towards a region of higher stability
  for the FQHE (see, e.g., Fig. 5). \\

\section{Discussion and conclusions}

Consider first a bilayer electron system with inter-layer separation $d$ but with zero inter-layer tunneling ($\alpha=0$). If $d/l_B$ is very large, there should be no FQHE at $\nu=1/2$ as the electron system behaves as two
independent layers with little or no inter-layer interaction, each at
filling factor $\nu=1/4$. As the layers are brought closer together
($d/l_B \lesssim 2)$, a $\nu=1/2$ FQHE, described by the two-component
$\Psi_{331}$ state, is possible. This is indeed believed to be the
case for the $\nu=1/2$ FQHE observed in bilayer electron systems in
double QW samples with very little inter-layer tunneling
\cite{Eisenstein.PRL.1992, He.PRB.1993, Note6}.

% \footnote{It is worth
%   emphasizing that the experimental phase diagram we present in Fig. 7
%   is for electron systems confined to a \textit{single wide} GaAs
%   QWs. The data of Ref. \cite{Eisenstein.PRL.1992}, which exhibit
%   $\nu=1/2$ FQHE in \textit{double} GaAs QWs with very small
%   inter-layer tunneling ($\alpha\simeq 0.01$) and $d/l_B\simeq 2$,
%   fall well outside the yellow-shaded region of Fig. 7 where the 1/2
%   FQHE is stable in wide GaAs QWs. An intriguing question is how far
%   would the yellow region in Fig. 6 extend for even wider GaAs QWs,
%   and whether it would ever contain the parameter range where the 1/2
%   FQHE is stable in double QWs.}

For finite values of tunneling
($\alpha>0$) two types of FQHE states can exist at $\nu=1/2$. At small
values of $\alpha$ the intra-layer correlations can dwarf the
tunneling energy, thus stabilizing a two-component $\Psi_{331}$ FQHE
state. For larger $\alpha$, on the other hand, the one-component
(Pfaffian) FQHE might be stable. For very large $\alpha$, it is most
likely that the ground state is compressible. The relative stabilities
of a $\nu=1/2$ $\Psi_{331}$ or Pfaffian FQHE state compared to a
compressible state are of course also influenced by the ratio of the inter-layer to
intra-layer interactions ($d/l_B$) as well as the thickness of the
"layers" (the parameter $\lambda$ in the insets to Figs. 1 and 7) which, as we discuss below,
softens the intra-layer short-range interactions
\cite{Suen.PRL.1992, Suen.PRL.1992b, Suen.Thesis.1993, Suen.PRL.1994}.

Following some of the earlier theoretical work \cite{Rezayi.BAPS.1987,
  Chakraborty.PRL.1987, Yoshioka.PRB.1989, MacDonald.SS.1990,
  Moore.Nuc.Phy.1991, He.PRB.1991, Greiter.PRL.1991, He.PRB.1993,
  Halperin.SS.1994}, recently there have been several reports of
calculations to determine the stability of the different FQHE and
compressible states in double and wide QWs \cite{Nomura.JPSJ.2004,
  Papic.PRB.2009a, Papic.PRB.2009, Storni.PRL.2010, Peterson.PRB.2010,
  Peterson.PRB.2010b}. A main conclusion of these calculations is that
the Pfaffian FQHE state at $\nu=1/2$ is not stable in a strictly 2D
electron system with zero layer thickness \cite{Storni.PRL.2010}. However, for a system with
finite (non-zero) layer thickness and tunneling, such as electrons in
a wide QW, the calculations indicate that at $\nu=1/2$ the Pfaffian
FQHE state could in principle exist at relatively large values of
$\alpha$. In particular, the $\Psi_{331}$ FQHE state is found to be
stable in a range of small values of $\alpha$ but as $\alpha$
increases the system makes a transition into a Pfaffian state before
it becomes compressible at very large values of $\alpha$
\cite{Nomura.JPSJ.2004, Peterson.PRB.2010}. Interestingly, the
calculations suggest that very near the transition between the
$\Psi_{331}$ and Pfaffian states, the energy gap for the 1/2 FQHE is
maximum, i.e., there is an upward cusp in the energy gap vs $\alpha$
plot \cite{Nomura.JPSJ.2004, Peterson.PRB.2010}. This is indeed what
is qualitatively seen in experiments \cite{Suen.Thesis.1993,
  Suen.PRL.1994}. In Ref. \cite{Nomura.JPSJ.2004} it was concluded
that the presence of this cusp and the finite value of the energy gap
on the Pfaffian side suggest that the Pfaffian state could be stable
at $\nu=1/2$ in wide QWs. Authors of Ref.~\cite{Peterson.PRB.2010}, on
the other hand, concluded that although a Pfaffian FQHE state at $\nu$
= 1/2 could in principle exist near this cusp, the chances that such a
state would survive in the thermodynamic limit are slim
\cite{Storni.PRL.2010}.

The phase diagrams we present here provide the comprehensive experimental conditions under which the $\nu=1/2$ FQHE
is observed in wide GaAs QWs with symmetric charge distributions. We note that the samples whose data we summarize in these diagrams span a large parameter range (well width, density, mobility), and yet the observed trends are remarkably consistent among the different samples. Of special interest are the samples with relatively narrow QW width, i.e., with large $\alpha$ where the Pfaffian state is most likely to exist.

It is worth closely comparing our Fig. 7 diagram with the latest theoretical diagrams, namely those in Fig. 5 of Ref. \cite{Peterson.PRB.2010}. Suppose we assume that the $\nu=1/2$ FQHE we observe is a $\Psi_{331}$ state and associate the dashed curve in Fig. 7 which separates the FQHE from the compressible state with the dashed curve in Fig. 5 of Ref. \cite{Peterson.PRB.2010} which separates the $\Psi_{331}$ FQHE state from the (unstable) Pfaffian state. There is overall some qualitative agreement between the characteristics of these curves in that they both have positive slopes and cover approximately the same ranges of $\alpha$ and $d/l_B$. There are, however, major quantitative discrepancies. For example,
the experimental boundary in Fig. 7 is relatively horizontal and
suggests that the critical $d/l_B$ values for the transition are
between $\simeq$ 5 and 6 in the range of $0.05<\alpha<0.16$ \cite{Note8}.
%\footnote{It is worth noting that in our samples the value of $W/l_B$ at the
%  compressible-FQHE boundary is also nearly a constant and ranges
%  between 8.0 and 9.5.}.
The boundary calculated in Ref. \cite{Peterson.PRB.2010}, on the other hand, has a much stronger dependence on $\alpha$. Moreover, in the phase diagram of Fig. 7, $\alpha$ for
our narrowest QW ($W=41$ nm) in which we observe the $\nu=1/2$ FQHE is
$\simeq$ 0.16. This $\alpha$, which is the largest value at which a FQHE at
$\nu=1/2$ has been observed, exceeds typical values of $\alpha$
where the $\Psi_{331}$ is theoretically predicted to be stable
\cite{Peterson.PRB.2010}. Interestingly, the plots in Fig. 5 of Ref. \cite{Peterson.PRB.2010} imply that the $\Psi_{331}$ FQHE state would be stable at $\alpha=0.15$ only when the bilayer system has a very narrow layer thickness (measured in units of $l_B$). However, our data summarized in Fig. 7 upper-right inset indicate that, for the FQHE we observe at $\alpha=0.16$ in the 41-nm-wide QW, the layer thickness $\lambda$ exceeds 3 times $\l_B$.

Despite the above discrepancies between the experimental and theoretical phase diagrams, we believe that the $\nu=1/2$ FQHE we observe in symmetric, wide GaAs QWs is very likely the $\Psi_{331}$ state. We base our conclusion on two experimental observations. First, as seen in Fig. 4, the $\nu=1/2$ FQHE is most stable when the charge distribution is symmetric and quickly disappears and turns into a compressible state when the distribution is made asymmetric \cite{Note7, Scarola.PRB.2010}. Similar observations were reported for wider GaAs QWs \cite{Suen.Thesis.1993,
  Suen.PRL.1994, Manoharan.PRL.1996, Shayegan.SST.1996, Shayegan.TALDS.1999, Shabani.PRL.2009a}. Our data of Fig. 4 demonstrate that this trend of extreme sensitivity to charge distribution symmetry extends to the narrowest QWs which have the largest values of $\alpha$.  Concurring with the conclusion reached in these previous
experimental studies \cite{Suen.Thesis.1993, Suen.PRL.1994,
  Manoharan.PRL.1996, Shayegan.SST.1996, Shayegan.TALDS.1999, Shabani.PRL.2009a}, we
suggest that the $\nu=1/2$ FQHE observed in symmetric, wide GaAs QWs
is a two-component state, stabilized by a delicate balance between the
intra-layer and inter-layer correlations.

Second, an examination of the parameters where we see the $\nu=1/2$ FQHE favors the above conjecture. Note that} the two-component $\Psi_{331}$ state is theoretically expected to be stable when the intra-layer and
inter-layer Coulomb interactions are comparable, i.e., have a ratio of
about unity \cite{Yoshioka.PRB.1989, He.PRB.1991}. For an ideal
bilayer electron system (with zero layer thickness), the ratio $d/l_B$
accurately reflects the relative strengths of the intra-layer and
inter-layer Coulomb interactions and the $\nu=1/2$ FQHE should be
observable for $d/l_B\lesssim 2$. However, in an electron system whose
layer thickness is comparable to or larger than $l_B$, the short-range
component of the Coulomb interaction, which is responsible for the
FQHE, softens \cite{Shayegan.PRL.1990, He.PRB.1990}. This softening is
significant in bilayer electron systems confined to wide QWs and, in
particular, $\lambda/l_B$ (for each layer) is about 2.4 to 3.3 in the
regime where the $\nu=1/2$ FQHE is stable (see Fig. 7 upper-right
inset) \cite{Suen.PRL.1992, Suen.PRL.1992b, Suen.Thesis.1993, Suen.PRL.1994}. Associating the $\nu=1/2$ FQHE we observe with the
$\Psi_{331}$ state, it is thus not surprising that we see this FQHE
when $d/l_B$ is much larger than unity (Fig. 7): The short-range component of
the intra-layer interaction is weaker for a bilayer system with larger
$\lambda/l_B$; therefore to ensure the proper intra-layer to
inter-layer interaction ratio which favors the $\Psi_{331}$ state, a
relatively weaker inter-layer interaction (smaller $e^2/ 4 \pi
\epsilon d$) is also needed, implying a larger $d/l_B$
\cite{Suen.PRL.1992, Suen.PRL.1992b, Suen.Thesis.1993, Suen.PRL.1994}. It is clearly evident in Fig. 7 upper-right inset that for a
sample with larger $\lambda/l_B$, the $\nu=1/2$ FQHE is indeed stable in a
region of larger $d/l_B$. \\

\section{Summary}

The phase diagrams we present here establish the experimental conditions under which the $\nu=1/2$ FQHE is observed in electron systems confined to symmetric, wide GaAs QWs in large ranges of density and QW width. The disappearance of the FQHE when the charge distribution is made asymmetric (Fig. 4) suggests that the $\nu=1/2$ FQHE has a two-component origin even in narrow QWs where the inter-layer tunneling is quite significant. We also compare our experimental phase diagram of normalized inter-layer distance vs tunneling (Fig. 7) to the theoretical diagrams which have been used to conclude a two-component origin for the $\nu=1/2$ FQHE. We find some agreement between the experimental and theoretical phase diagrams, but there are certainly quantitative discrepancies. Our observation that in a simple tunneling vs density diagram (Fig. 5) the boundary between the FQHE and compressible state is essentially a
straight line is also intriguing and seeks explanation. We
conclude that, in light of our new data, more detailed and precise
calculations are needed to make a quantitative connection to the experimental data and help unveil the physics of the $\nu=1/2$ FQHE in
symmetric, wide GaAs QWs.\\

We acknowledge support through the NSF (Grants DMR-0904117, DMR-1305691, and MRSEC DMR-0819860), the Keck Foundation, and the Gordon and Betty Moore Foundation (Grant GBMF2719). A portion of this work was performed at the National High Magnetic Field Laboratory which is supported by National Science Foundation Cooperative Agreement No. DMR-1157490, the State of Florida and the US Department of Energy. We are grateful to J.K. Jain and Z. Papic for
illuminating discussions, and to E. Palm, S. Hannahs, T. P. Murphy,
J. H. Park and G. E. Jones for technical assistance with the high magnetic field measurements.

\bibliography{../bib_full}

\end{document}